\documentclass[10pt,letterpaper]{article}
\usepackage{opex3}
\usepackage{amsmath}
\usepackage{bm}

\begin{document}
\newcommand{\bq}{\begin{subequations}}
\newcommand{\eq}{\end{subequations}}
\newcommand{\beq}{\begin{eqnarray}}
\newcommand{\eeq}{\end{eqnarray}}

\newcommand{\st}[1]{{\mbox{${\mbox{\scriptsize #1}}$}}}  
\newcommand{\BM}[1]{\mbox{\boldmath $#1$}}              
\newcommand{\uv}[1]{\mbox{$\widehat{\mbox{\boldmath $#1$}}$}}   
\newcommand{\dy}[1]{\mbox{\boldmath $\overline{#1}$}}   
\newcommand{\sbtex}[1]{{\mbox{\scriptsize #1}}}   
\newcommand{\nab}{\mbox{\boldmath $\nabla$}}            
\newcommand{\CS}{\mbox{$\begin{array}{c}\cos \\ \sin \\ \end{array}$}}
\newcommand{\SC}{\mbox{$\begin{array}{c}\sin \\ \cos \\ \end{array}$}}
\def\centertiff#1#2#3{\vskip#2\relax\centerline{\hbox to#1{\special
  {bmp:#3 x=#1, y=#2}\hfil}}}
\def\twoeps#1#2#3#4{\vskip#2\relax\centerline{{\hbox to#1{\special
  {eps:#3 x=#1, y=#2}\hfil}}\hspace{-1ex}
  {\hbox to#1{\special {eps:#4 x=#1, y=#2}\hfil}}}}

\newcommand{\citeasnoun}[1]{Ref.~\citenum{#1}}

\def\a{s}
\def\b{s}
\newcommand{\add}[1]{\if\a\b{{\color{red} #1}}\else{#1}\fi}
\newcommand{\comm}[1]{\if\a\b{\marginpar{\color{blue}\{\small #1\}}}\else{}\fi}
\newcommand{\del}[1]{{\if\a\b{{\color{magenta}[[#1]]}}\else{}\fi}}

\title{Spherical cloaking using nonlinear transformations for improved segmentation into concentric isotropic coatings}
\author{Cheng-Wei Qiu$^{1,2}$, Li Hu$^{2}$, Baile Zhang$^{1}$, Bae-Ian Wu$^{1}$, Steven G. Johnson$^{3}$, and John D.
Joannopoulos$^{1}$}
\address{$^{1}$Research Laboratory of Electronics, Massachusetts Institute of Technology, Cambridge, Massachusetts 02139, USA}\email{cwq@mit.edu} 
\address{$^{2}$Department of Electrical and Computer Engineering, National University of Singapore, 4 Engineering Drive 3, Singapore 117576.}
\address{$^{3}$Department of Mathematics, Massachusetts Institute of Technology, Cambridge, Massachusetts 02139, USA}\email{stevenj@math.mit.edu}


\begin{abstract}
Two novel classes of spherical invisibility cloaks based on
nonlinear transformation have been studied. The cloaking
characteristics are presented by segmenting the nonlinear
transformation based spherical cloak into concentric isotropic
homogeneous coatings. Detailed investigations of the optimal
discretization (e.g., thickness control of each layer, nonlinear
factor, etc.) are presented for both linear and
nonlinear spherical cloaks and their effects on invisibility
performance are also discussed. The cloaking properties and our
choice of optimal segmentation are verified by the numerical
simulation of not only near-field electric-field distribution but
also the far-field radar cross section (RCS).
\end{abstract}



\section{Introduction}
The use of coordinate transformation to control electromagnetic fields
\cite{Pendry_sci} has been receiving extensive attention
\cite{Schurig_OE2006,Leonhardt_sci2006,Schurig_OE2007}. This approach
was generalized from the cloaking of thermal conductivity
\cite{Greenleaf_2003}, and was further widely applied in
electromagnetics and acoustics \cite{Pendry_sci,CTChan_2007}, which
provides new ways to conceal passive/active objects \cite{
  ChenHS_2007,ZhangBL_2008} within their interiors invisible to
external illuminations. The fundamental idea is that Maxwell's
equations are invariant under a coordinate transformation if the
material properties (electric permittivity and magnetic
permeability) are altered appropriately; i.e., a specific spatial
compression is equivalent to a variation of the material parameters
in flat (original) space. Cloaks in regular shapes (e.g., spherical
\cite{Pendry_sci}, square \cite{square}, cylindrical
\cite{cylinder}, or elliptic \cite{ellipse}) have been proposed
based on the construction of explicit transformation matrices. There
are many works devoted to the investigation of 2D
cylindrical/elliptic cloaks approximated by, for example, simplified
material parameters or equations
\cite{simplified1,simplified2,equation,optical}, or designed for
limited incident angles \cite{limitation}. Experimental
investigations of 2D cloaks have demonstrated significant reductions
in the scattering cross section, albeit for narrow bandwidths in the
microwave regime and for objects at most a few wavelengths in
diameter so far \cite{microwave}. Broader bandwidths have been
demonstrated for cloaking an object on a ground plane, but again
only for a few wavelengths diameter~\cite{groundplane}. To design
irregularly shaped cloaks, both analytical \cite{theoretical1} and
numerical methods \cite{numerical1,numerical2} have been proposed.
In what follows, we will confine our discussion within the area of
3D electromagnetic spherical cloaks.

Inspired by the classic spherical cloak \cite{Pendry_sci} based
  on a linear coordinate transformation, the expressions of
  electromagnetic fields were explicitly presented in terms of
  spherical Bessel functions along the lines of Mie scattering theory
  \cite{ChenHS_2007}, and it again was demonstrated that the external
  incident wave cannot interact with the cloaked object in a perfect
  spherical cloak. Linear
coordinate transformations correspond to a transverse/radial
permittivity ratio (\emph{anisotropy ratio} \cite{Qiu_PRE2007})
$\epsilon_t/\epsilon_r=r^2/(r-R_1)^2$, where $R_1$ is the radius of
the cloaked region. The subtle point of the singularity in the
  coordinate transformation at the innner surface of the cloak was
  analytically shown to correpond to surface voltages
  \cite{ZhangBL_2008} in spherical cloaks, leading to an explicit
  physical explanation for why the wave cannot leave the cloaked
  region and external waves cannot enter that region. These works on
  spherical cloaks considered linear coordinate transformations, while
  more recently nonlinear (``high-order'') transformations have been
  considered in order to obtain more degrees of freedom in designing
  the material parameters \cite{high-order1,high-order2}. In this paper, we consider the
  utility of nonlinear coordinate transformations to improve the
  performance of the cloak when it is approximated (segmented) into a
  sequence of piecewise-homogeneous layers.  If the anisotropy ratio
differs from $r^2/(r-R_1)^2$, it will lead to quite complicated
formulations for the field expressions, which cannot be treated by the
previous method for a position-dependent anisotropy ratio \cite{ChenHS_2007} or
by the method for a constant anisotropy ratio \cite{Qiu_AWPL,Qiu_TAP07}.
Here, in order to study the proposed nonlinear-transformation
spherical cloaks, each anisotropic cloaking shell is discretized
into multiple alternating homogeneous isotropic coatings, not only for
the ease of fabrication but also for the computation convenience in dealing with
arbitrary anisotropy ratios.

In general, practical implemenations of passive invisibility
  cloaks by coordinate transformation are limited by several factors,
  in addition to the bandwidth limitations (from material dispersion)
  mentioned above.  First, there are fabrication imperfections and the
  finite size of the subwavelength components of the metamaterials.
  Second, there is material absorption, which is especially
  challenging when an extreme resonant response is required of the
  material in order to obtain very large or very small
  permittivity/permeability. Third, there is the difficulty of
  fabricating continuously varying anisotropic materials.  In this
  paper, we focus only on the third issue. A common approximation for
  continuously varying materials is to segment them into
  piecewise-homogeneous layers, and we show that the choice of
  coordination transformation has a significant impact upon the
  success of this approximation. We argue that the segmentation is
  best performed in virtual space rather than in the physical
  (transformed) coordinates. We also demonstrate that the anisotropic
  materials can be further approximated by a sequence of
  isotropic-material layers. The optimal selection of material
parameters, nonlinear transformation, and segmentation (in virtual
space), in order to achieve a low scattering cross sectikon over a
wide range of observation angles, is found for nonlinear spherical
cloaks and verified by numerical results.

\section{Preliminaries}
\begin{figure}[htbp]
\centering\includegraphics[width=5.4cm]{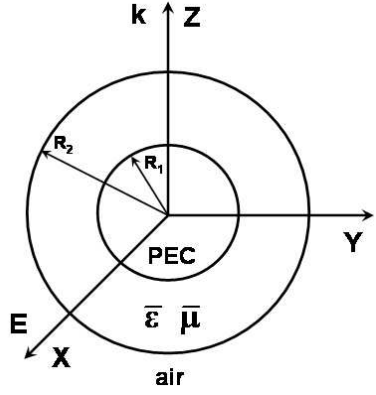} \caption{The geometry
of a spherical cloaking structure.}\label{Geometry}
\end{figure}

The configuration of the spherical cloaking structure is shown in
Fig.~\ref{Geometry}. The inner region ($r<R_1$) is a perfect
electric conductor (PEC) and the intermediate region ($R_1<r<R_2$)
is filled by the nonlinearly transformed spherical cloak, characterized by
$\dy{\epsilon}(r)$ and $\dy{\mu}(r)$, which will be discussed below.
The electric field is polarized along the $x$~axis and propagating along
the $z$~axis.

By applying the decomposition method and separation of variables
\cite{Qiu_PRE2007}, the equation for the radial component of the field
potentials becomes \beq\label{EqR(r)} \Big\{\frac{\partial^2}{\partial
  r^2}+\Big[k^2_t-\mbox{A}_{e,m}\frac{n(n+1)}{r^2}\Big]\Big\}\mbox{f}(r)=0,
\eeq where the subscripts $e$ and $m$ denote the electric and magnetic
anisotropic ratio, respectively, and
$k_t=\omega\sqrt{\mu_t\epsilon_t}$ \cite{Qiu_PRE2007}. For
$\dy{\bm{\epsilon}}$ and $\dy{\bm{\mu}}$ of perfect linear spherical
cloaks \cite{Pendry_sci}, i.e., $A_e=A_m=r^2/(r-R_1)^2$,
Eq.~(\ref{EqR(r)}) is reduced to \beq\label{EqR(r)-linear}
\Big\{\frac{\partial^2}{\partial
  r^2}+\Big[k^2_t-\frac{n(n+1)}{(r-R_1)^2}\Big]\Big\}\mbox{f}(r)=0.
\eeq Thus, the radial component $f(r)$ can be solved in a way
similar to that for isotropic materials, except for the change in
the argument of resultant Bessel/Hankel functions. However, given a
set of $\dy{\bm{\epsilon}}$ and $\dy{\bm{\mu}}$ derived from a
certain transform, the anisotropy ratio may not be $r^2/(r-R_1)^2$
anymore, and then the radial component cannot be solved explicitly
in the same way. In this situation, we could approximate the
original inhomogeneous anisotropic cloak by many thin, homogeneous
anisotropic coatings, and the diffraction problem can thus be solved
in terms
  of analytical Bessel/Hankel functions satisfying boundary conditions
  at each interface. Nevertheless, the requirement of anisotropic
materials still remains. Alternatively, to further alleviate the
restriction in material complexity, the original inhomogeneous
anisotropic cloaking materials can be approximated by the limit
  of many thin, concentric, homogeneous, \emph{isotropic} coatings, forming an effective anisotropic medium, which will be discussed in
this section.

\begin{figure}[htbp]
\centering\includegraphics[width=12.5cm]{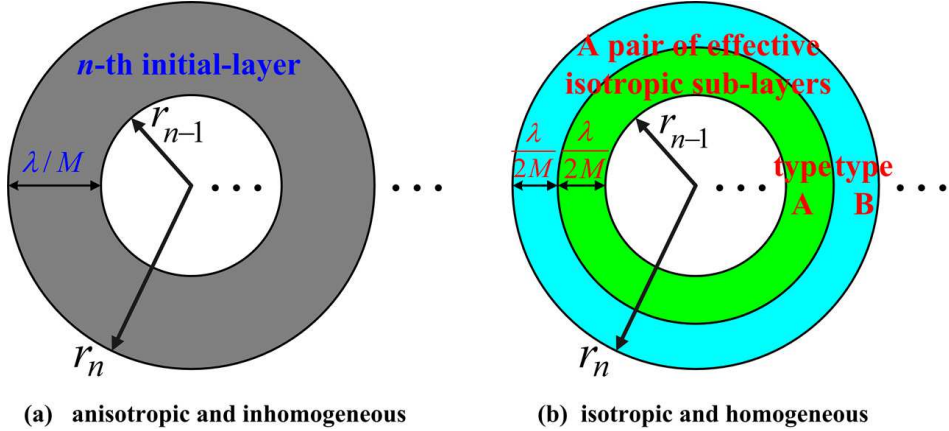} \caption{The
discretization of a general anisotropic spherical cloak (a) into an
equivalent isotropic coated structure (b). An illustrative example
for the conversion from the $n$-th anisotropic initial-layer (width
is $r_n-r_{n-1}$) into its two sub-layers (isotropic) with equal
thickness $(r_n-r_{n-1})/2$ has been shown.}\label{Conversion}
\end{figure}

First, we consider a general spherical cloak characterized by
$\dy{\bm{\epsilon}}=\epsilon_r(r)\uv{r}\uv{r}+\epsilon_t(r)(\uv{\theta}\uv{\theta}+\uv{\phi}\uv{\phi})$
and
$\dy{\bm{\mu}}=\mu_r(r)\uv{r}\uv{r}+\mu_t(r)(\uv{\theta}\uv{\theta}+\uv{\phi}\uv{\phi})$,
in which $\epsilon_r(r)=\mu_r(r)$ and $\epsilon_t(r)=\mu_t(r)$ are
position-dependent in general. It is then divided into $M$
initial-layers (anisotropic) but the thickness of individual initial
layers may or may not be identical, depending on the transformation.
Given a coordinate transformation function $r'=f(r)$ between the
virtual space (i.e., $\Omega'(r')$, $0<r'<R_2$) and the compressed
space (i.e., $\Omega(r)$, $R_1<r<R_2$), the stepwise segmentation in
physical space ($r$) is desired to mimic the transformation function
as well as possible and we also desire that the segmentation not be
too complicated. We find that equally dividing the \emph{virtual}
space ($r'$) into $M$ initial-layers will make the segmentation on
the physical space ($r$) ``self-adaptive'' in a simple way---it
automatically uses a finer
  discretization in regions of physical space where the anisotropic
  materials are varying more rapidly---which will result in better
invisibility performance.  This conversion is illustrated in
Fig.~\ref{Conversion}. Throughout this paper, we apply the same
condition, i.e., the segmented layers in $r'$ are of equal
thickness, which represents a good choice of segmentation in the
initial-layers in the cloak shell $R_1<r<R_2$. By projecting the
segmentation in $r'_n=R_2\cdot n/M$ onto the physical $r$, one has
$r_n=f^{-1}(r'_n),~(\mbox{n=1,~2,...,~M})$. Thus, the geometry of
every initial-layer in Fig.~\ref{Conversion}(a) is determined.


Subsequently, we mimic each spherically anisotropic initial layer by
a pair of effective isotropic sub-layers (type-A and type-B) with
equal thickness at $(r_n-r_{n-1})/2$, as shown in
Fig.~\ref{Conversion}(b). The material parameters of type-A and
type-B isotropic dielectrics can be inferred and derived from the
result for radial conductivity by Sten \cite{Sten}:
\bq\label{condition}\beq
&&\epsilon_A=\mu_A=\epsilon_t+\sqrt{\epsilon^2_t-\epsilon_t\epsilon_r}\\
&&\epsilon_B=\mu_B=\epsilon_t-\sqrt{\epsilon^2_t-\epsilon_t\epsilon_r}.
\eeq\eq Now, the original spherical cloak turns to be a concentric
isotropic coatings. Throughout this paper, $M=40$ and
$R_2=2R_1=2\lambda$ are chosen in order that the validity of
Eq.~(\ref{condition}) can be maintained, i.e., sub-layers have to be
sufficiently thin. Alternatively one could also choose a smaller $M$
associated with the corresponding parameters determined by
optimization methods and then subdivide each anisotropic layer into
many more than two isotropic sub-layers.

When converting the $n$-th initial layer into its pair of effective
isotropic sub-layers, we need to pick a specific radial position for
those $\epsilon_r$ and $\epsilon_t$ on the right-hand side of
Eq.~(\ref{condition}) within the initial layer in order to determine
corresponding parameters of the two isotropic dielectrics
($\epsilon_A$,~$\mu_A$) and ($\epsilon_B$,~$\mu_B$) on the left-hand
side of Eq.~(\ref{condition}) and in Fig.~\ref{Conversion}(b).
According to \cite{Tradeoff}, such a discretization mechanism using
$r=r_n$ in Eq.~(\ref{condition}) for the $n$-th initial layer will
give a good compromise for both forward and backward scatterings,
while retaining good invisibility performance.

\section{Effects of Nonlinear Transformation in Nonlinear Spherical Cloaks}
Pendry's spherical cloak \cite{Pendry_sci} is a linear one, and
hence it is a straight line if one plots the transformation function
$r'$ against the physical radius $r$. In what follows, we introduce
two classes of nonlinear-transformation spherical cloaks, and we
discuss how to restore and improve the invisibility performance
after discretization by choosing a proper nonlinear transformation
and by choosing a suitable compensation scheme while discretizing
the original spherical cloak into multilayer isotropic structures.
The two types of spherical cloaks considered here are classified in
terms of the negative (i.e., concave-down) or positive (i.e.,
concave-up) sign of the second derivative of the transformation
function. It is important to reiterate that all three
  designs---linear, concave-up, and concave-down---are perfect cloaks
  for the exact inhomogeous design, and we are only considering the
  breakdown of invisibility when the design is discretized into
  homogeneous layers.

\subsection{Concave-Down Nonlinear Transformation}
To compress a sphere of air at the radius $R_2$ in $\Omega'$
(original) space into a shell at the region $R_1<r<R_2$ in $\Omega$
(compressed) space, we propose a class of prescribed functions
\beq\label{function}
r'(x)=\frac{R_2^{x+1}}{R_2^x-R_1^x}[1-(\frac{R_1}{r})^x] \eeq where
$x$ denotes the degree of the nonlinearity in the transformation.
When $x$ is very small in Eq.~(\ref{function}), the curves
are difficult to distinguish and all approach to the same limiting case
when $x\rightarrow 0$:
\begin{equation}\label{limiting}
r'(x)=\frac{R_2\mbox{Ln}[r/R_1]}{\mbox{Ln}[R_2/R_1]}.
\end{equation}

Thus the parameters $\dy{\bm{\epsilon}},~\dy{\bm{\mu}}$ in the
transformed coordinates can be written in term of
$\dy{\bm{\epsilon}}',~\dy{\bm{\mu}}'$ in the original space by
\beq\label{newParam} \dy{\bm{\epsilon}}(r)=
A\dy{\bm{\epsilon}}'(r')A^T/det(A),~~~~\dy{\bm{\mu}}(r)=
A\dy{\bm{\mu}}'(r')A^T/det(A)\eeq where $A$ is the Jacobian matrix
with elements defined as $A_{ki}=\partial r_k/\partial r'_i$.

\begin{figure}[htbp]
\centering\includegraphics[width=8.1cm]{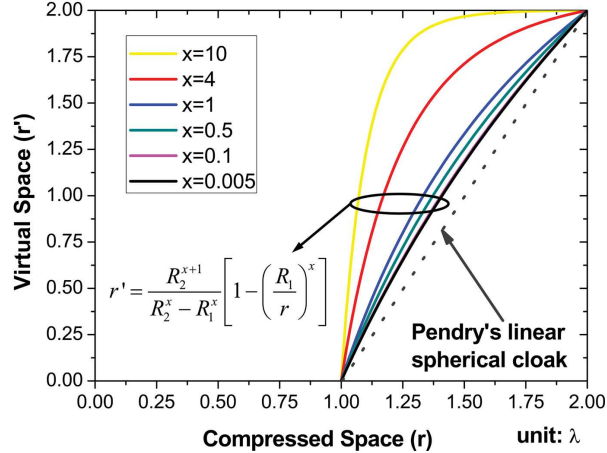}
\caption{The concave-down nonlinear transformation. When $x<0.1$,
all mapping curves are overlapping, meaning they effectively lead to
the same performance. When $x$ becomes large, the curve will nearly
become a step function over
$1\lambda<r<2\lambda$.}\label{Upper-Class}
\end{figure}

Note that all curves belonging to those transformation functions in
Eq.~(\ref{function}) have negative second derivative with respect to
the physical space $r$, and we term this class of transformations
the \emph{concave-down} nonlinear transformation. The nonlinear
transformation function in Eq.~(\ref{function}) only depends on the
radial component $r$ in the spherical coordinate system
$(r,~\theta,~\phi)$. Thus it is easy to find that the Jacobian
matrix $A$ is diagonal. Considering that the original space is filled
with air ($\dy{\bm{\epsilon}}'=\dy{\bm{\mu}}'=1$),
Eq.~(\ref{newParam}) can be rewritten as \beq
\dy{\bm{\epsilon}}=\dy{\bm{\mu}}=
\mbox{diag}[\lambda^2_r,~\lambda^2_\theta,~\lambda^2_\phi]/\lambda_r\lambda_\theta\lambda_\phi
=\mbox{diag}[\frac{\lambda_r}{\lambda_\theta\lambda_\phi},~\frac{\lambda_\theta}{\lambda_r\lambda_\phi},~\frac{\lambda_\phi}{\lambda_r\lambda_\theta}]
\eeq where \bq \beq \lambda_r\hspace{-1.8ex}&=&\hspace{-1.8ex}\frac{\partial r}{\partial r'}=\frac{(R_2^x-R_1^x)r^{x+1}}{xR_1^xR_2^{x+1}}  \\
\lambda_\theta \hspace{-1.8ex}&=&\hspace{-1.8ex}
\lambda_\phi=\frac{r}{r'}=\frac{(R_2^x-R_1^x)r^{x+1}}{R_2^{x+1}(r^x-R_1^x)}
\eeq \eq denoting three principal stretches of the Jacobian matrix.
Finally, the desired parameters in
the compressed space ($R_1<r<R_2$) are shown to be \bq\label{finalParam} \beq \epsilon_r\hspace{-1.8ex}&=&\hspace{-1.8ex}\mu_r=\frac{R_2^{x+1}(r^x-R_1^x)^2}{xR_1^x(R_2^x-R_1^x)r^{x+1}} \\
\epsilon_\theta \hspace{-1.8ex}&=&\hspace{-1.8ex}
\mu_\theta=\epsilon_\phi=\mu_\phi=\frac{xR_1^xR_2^{x+1}}{(R_2^x-R_1^x)r^{x+1}}.
\eeq \eq Given such radial and transversal parameters, it is obvious
that Eq.~(\ref{EqR(r)}) cannot be simply solved due to the
complicated anisotropy ratio
$A_{e,m}=\epsilon_t/\epsilon_r=x^2R_1^{2x}/(r^x-R_1^x)^2$ in our
current case. In addition, one can also find that when the virtual
space is equally discretized using a concave-down nonlinear
transformation, the segmentation of the initial layers in
Fig.~\ref{Conversion} in the physical space $r$ is \beq
r_n(x)=\left[1-\frac{R_2^x-R_1^x}{R_2^x}\frac{n}{M}\right]^{-1/x}\cdot
R_1,~~~~~~~\mbox{n=1,~2,..., M}. \eeq

\begin{figure}[htbp]
\centering\includegraphics[width=11.1cm]{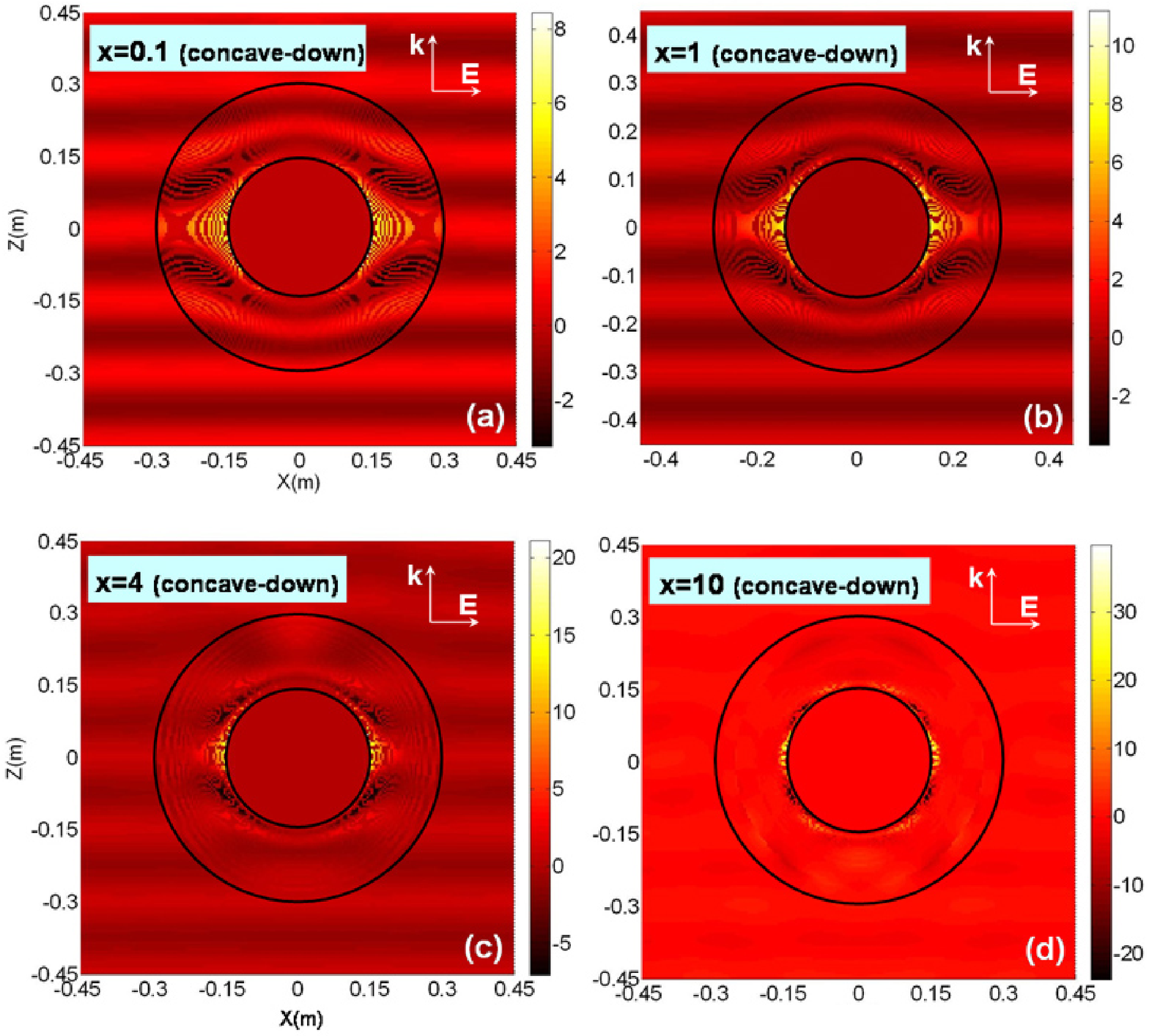}\\
\vspace{-2ex}{\small $\mbox{Re[E}_{\mbox{total}}\mbox{]}$ in all
regions on x-z plane} \vspace{2ex}\\
\centering\includegraphics[width=11.4cm]{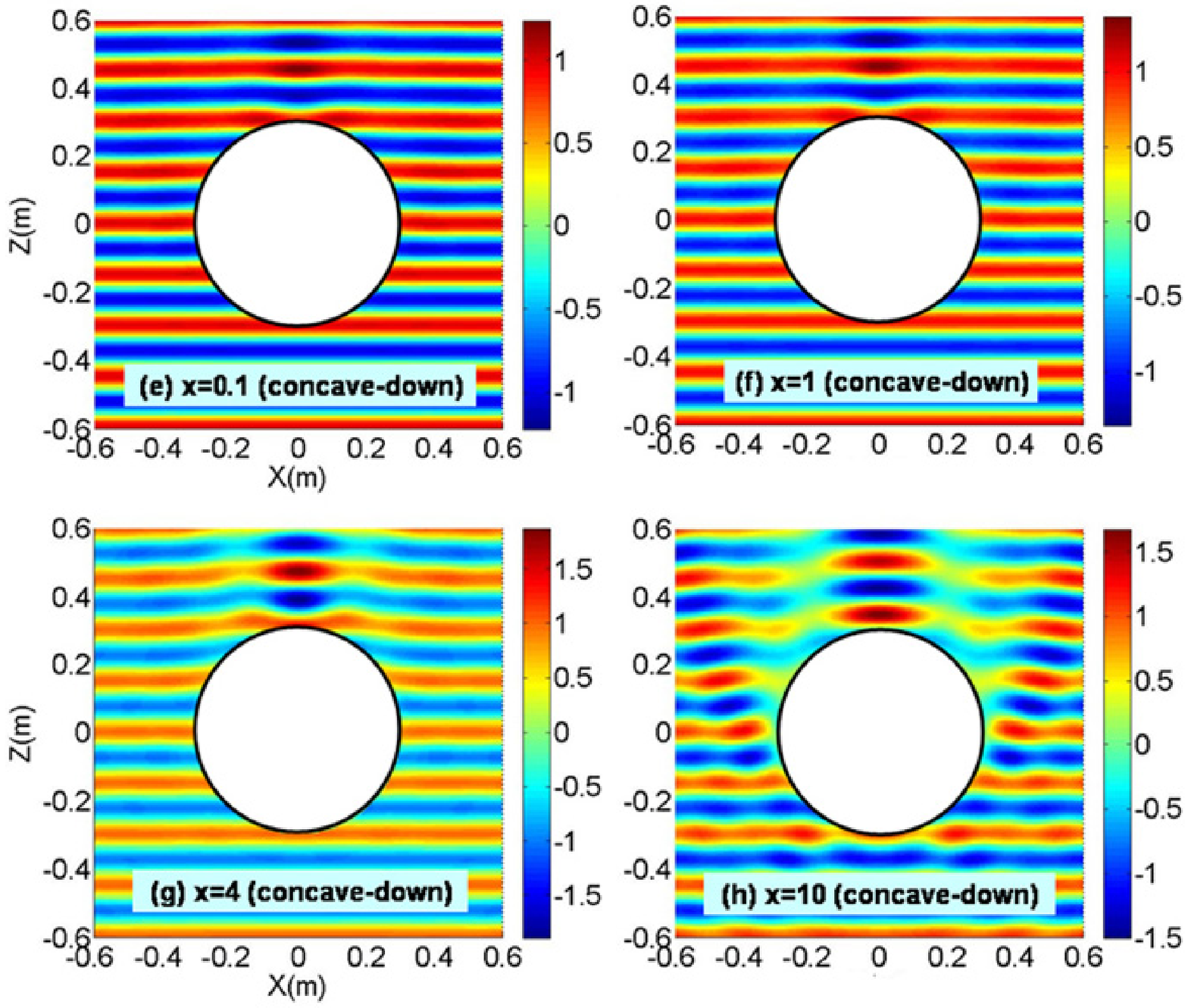}\\
\vspace{-2ex} {\small $\mbox{Re[E}_{\mbox{total}}\mbox{]}$ outside
the cloak on x-z
plane}\\
\caption{Near-field interaction in the presence of the proposed
nonlinear spherical cloak with different values of $x$ for the
concave-down class. The inner region $0<r<R_1$ is filled by PEC. $M$
is set to be 40. Frequency is 2GHz. The total electric fields are
plotted only in the region $r>R_2$ in (e-h).}\label{Upper-near}
\end{figure}


Here, we will study the effects of the nonlinear factor $x$ in the
near-field and far-field of the discretized nonlinear-transformation
spherical cloaks. We fix M=40, and consider the cases of x=0.1, x=1,
x=4, and x=10 in concave-down nonlinear cloaks, shown in
Fig.~\ref{Upper-near}. It can be seen from
Figs.~\ref{Upper-near}(a-d) that when $x$ increases, the magnitude
of electric field increases significantly inside the cloak. This is
because more energy is guided towards the inner boundary of the PEC
core, which in turn makes the cloaked PEC more {\itshape visible} to
external incidences. To prove that the large electric fields only
occur in the region $R_1<r<R_2$ and to study the effect of $x$ on
individual near-field perturbations more explicitly, only the fields
outside the cloak are presented in Figs.~\ref{Upper-near}(e-f). One
can again confirm that, for discretized concave-down nonlinear spherical cloaks,
smaller $x$ leads to smaller disturbance in electric fields in the
outer space. In particular, the degradation in the invisibility in
the forward direction is proportional to the value of $x$ (see
Fig.~\ref{Upper-near}(g) and Fig.~\ref{Upper-near}(h)).

\begin{figure}[htbp]
\centering\includegraphics[width=9.5cm]{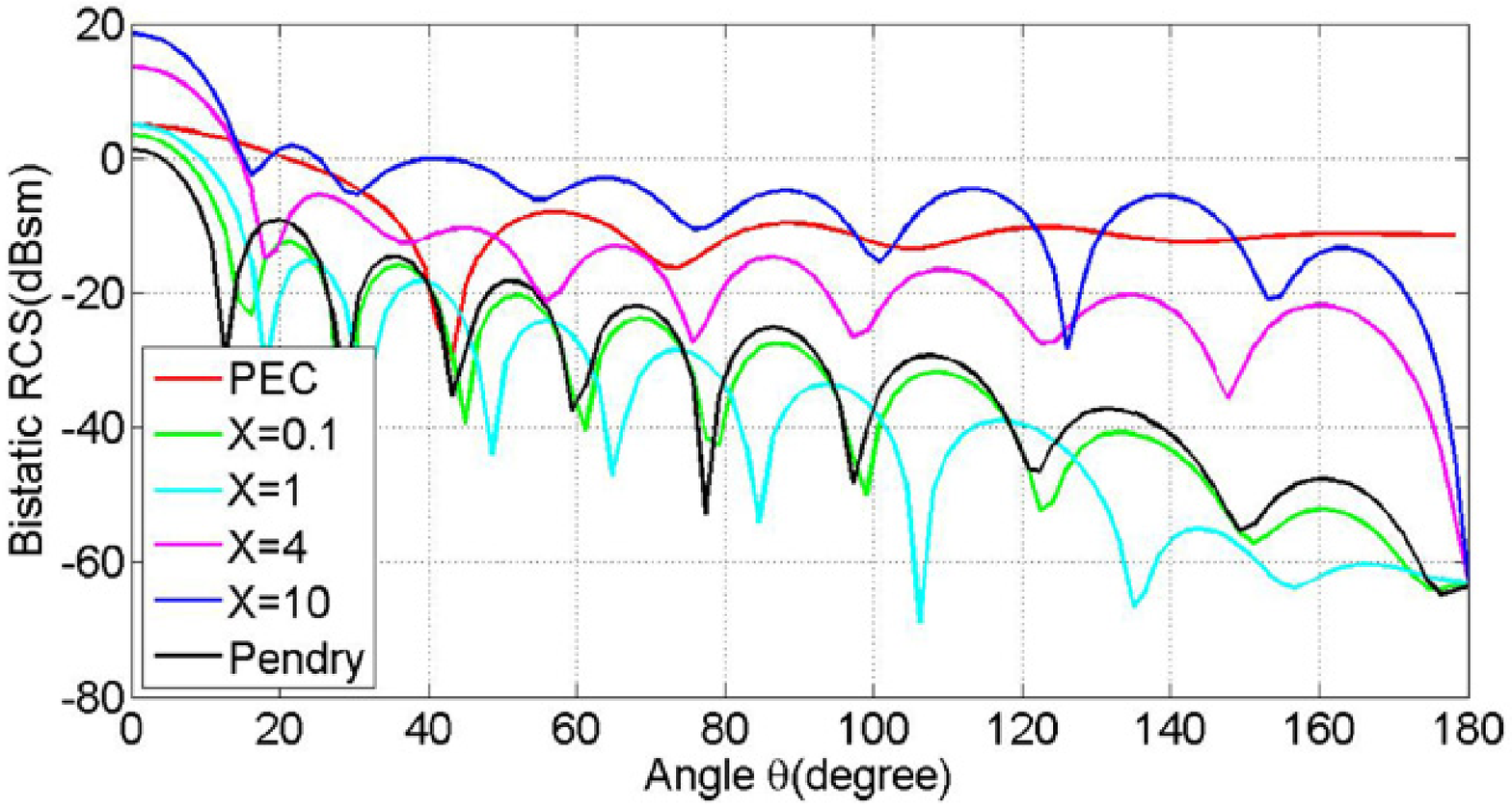}
\caption{Bistatic RCS of the concave-down nonlinear cloaks.
$M=40$.}\label{Upper-far}
\end{figure}

In Fig.~\ref{Upper-far}, we consider the far fields of those cases
of Fig.~\ref{Upper-near}. For comparison purposes, we introduce two
more curves, PEC and Pendry's cloak. It should be noted that
the curve corresponding to Pendry's cloak is a straight
line, and the same segmentation procedure as discussed above is applied to it, which
is the same as equally dividing it into $M$ initial-layers in the
physical space (see Fig.~\ref{Upper-Class}). From
Fig.~\ref{Upper-far}, it can be seen that for concave-down nonlinear
transformations, the cloaking property is better retained when $x$ is
small. Except for the forward direction, concave-down nonlinear
transformation with small $x$ can achieve even lower RCS over a wide
range of observation angles. As $x$ keeps increasing, their RCSs are
increasing dramatically and can be larger than that of a uncloaked
PEC core (e.g., $x=10$). No matter how large $x$ becomes, the
suppressed backward scattering is still maintained \cite{ChenHS_2007}
though it is not exactly zero due to the discretization, but
the forward scattering will become higher and higher, which is not
desired in the cloaking application. Those far-field characteristics
are consistent with the perturbation in corresponding near-field
patterns in Fig.~\ref{Upper-near}.

\subsection{Concave-Up Nonlinear Transformation}
We can also propose another class of prescribed functions
\beq\label{function-lower-class}
r'(x)=\frac{R_2R_1^{x}}{R_2^x-R_1^x}[(\frac{r}{R_1})^x-1]. \eeq It
is found that when $x\rightarrow 0$, it also approaches to the same
limit as Eq.~(\ref{limiting}). In the same manner, we can obtain the
desired parameters in
the compressed space ($R_1<r<R_2$) \bq\label{finalParam-lower-class} \beq \epsilon_r\hspace{-1.8ex}&=&\hspace{-1.8ex}\mu_r=\frac{R_2(r^x-R_1^x)^2}{x(R_2^x-R_1^x)r^{x+1}} \\
\epsilon_\theta \hspace{-1.8ex}&=&\hspace{-1.8ex}
\mu_\theta=\epsilon_\phi=\mu_\phi=\frac{xR_2 r^{x-1}}{R_2^x-R_1^x}.
\eeq \eq Note that the case $x=1$ is exactly Pendry's linear
spherical cloak.

\begin{figure}[htbp]
\centering\includegraphics[width=7.75cm]{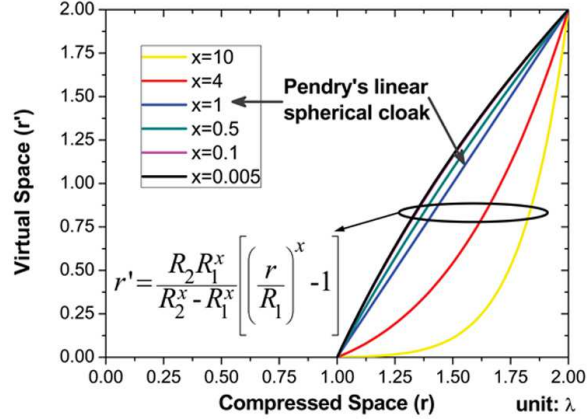}
\caption{The concave-up nonlinear transformation. When $x=1$, it is
exactly Pendry's linear cloak, which can be verified by inserting
$x=1$ into Eq.~(\ref{finalParam-lower-class}). When $x$ gets
extremely large, the mapping curve will nearly become a sharp
impulse when $r \rightarrow 2\lambda$.}\label{Lower-Class}
\end{figure}

It can be seen in Fig.~\ref{Lower-Class} that all curves belonging
to Eq.~(\ref{function-lower-class}) have positive second
derivatives, and we term this class of transformations as the
\emph{concave-up} nonlinear transformations. Also, it is found that the
limiting case of Eq.~(\ref{limiting}), i.e, $x\rightarrow 0$, is the
dividing line between the convace-down and concave-up classes. Since
$x \ll 1$ in Eq.~(\ref{function-lower-class}) is actually very close to
the situation of $x \ll 1$ for concave-down nonlinear transformations, we
will not revisit that case here. Instead, we only consider $x=1$, $x=4$, and
$x=10$ concave-up transformations [Eq.~(\ref{finalParam-lower-class})], and the segmentation for
concave-up nonlinear spherical cloaks in physical space $r$ turns out to
be \beq\label{lower-seg}
r_n(x)=\left[\frac{R_2^x-R_1^x}{R_1^x}\frac{n}{M}+1\right]^{1/x}\cdot
R_1,~~~~~~~\mbox{n=1,~2,..., M}. \eeq

\begin{figure}[htbp]
\centering\includegraphics[width=10cm]{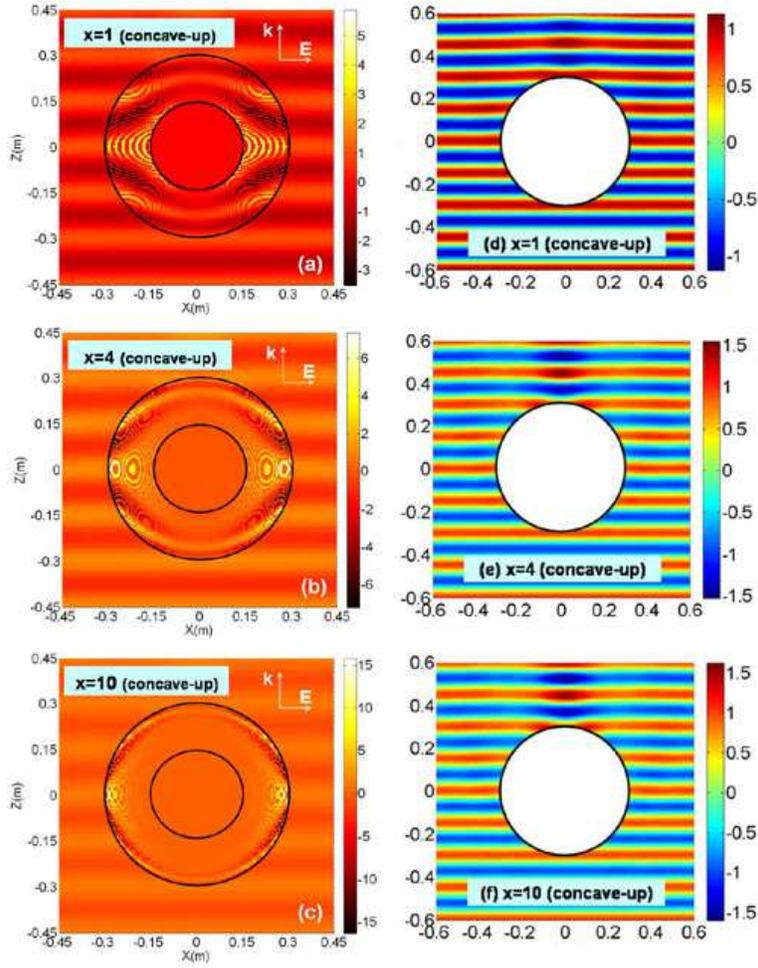}
\caption{$\mbox{Re[E}_{\mbox{total}}\mbox{]}$ on x-z plane for the
proposed concave-up nonlinear transformation cloaks at $x=1$, $x=4$
and $x=10$, respectively. Figs.~\ref{Lower-near}(a-c) present total
electric fields in all areas while Figs.~\ref{Lower-near}(e-f) show
the total fields only outside the cloak ($r>R_2$). $M=40$ and
$f=2GHz$. In order to show the disturbance in surrounding outer
space, the plot range is larger (from -0.6m to 0.6m) in the right
column.}\label{Lower-near}
\end{figure}

We are interested to see which class will give rise to better
invisibility performance, after discretization, if all the other conditions are the same.
We still plot the near field of concave-up cloaks at ``x=1'',
``x=4'', and ``x=10''. Comparing Fig.~\ref{Upper-near} with
Fig.~\ref{Lower-near}, one can see that: (1) the peak amplitude
of the electric field is proportional to $x$ in both classes; (2)
for each value of $x$, concave-up transformation cloaks have lower
peak amplitudes than concave-down transformation cloaks, i.e.,
the perturbation inside the cloak region is smaller; (3) the
perturbation outside the cloak of concave-up transformation cloaks
is also lower than that of concave-down transformation cloaks in all
cases of $x$; (4) the invisibility performance is better maintained for concave-up transformations even
when $x=10$ or even larger. Hence, spherical cloaks based on
concave-up nonlinear transformations exhibit better invisibility after discretization.

\begin{figure}[htbp]
\centering\includegraphics[width=9.1cm]{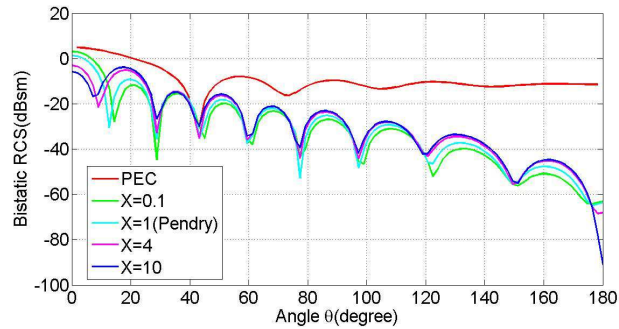}
\caption{Bistatic RCS of the concave-up nonlinear
cloaks.}\label{Lower-far}
\end{figure}

Fig.~\ref{Lower-far} exhibits some interesting features which
distinguish it from the far fields of concave-down transformations in
Fig.~\ref{Upper-far}. Over a wide range of angles, the invisibility
performance of spherical cloaks based on concave-up nonlinear
transformations are more stable under variations in $x$. More
importantly, when $x$ increases (e.g., $x=4$ or $x=10$), though the
RCS will be a bit larger than Pendry's linear cloak at most angles,
the RCS near the forward direction will be reduced greatly compared with
that of classic linear cloak. Such far-field phenomena are also
connected with the near-field patterns. It is important to point out
that the increase of $x$ in concave-up transformations [see
Figs.~\ref{Lower-near}(a--c)] will push the ``hot'' areas (where the
electric field is very high) further and further away from the
spherical PEC core, so the induced shadow and the far-field pattern
become stable with increasing $x$. However, the increase of $x$ in the
concave-down transformations has little effect in shifting away the
``hot" positions from the core [see Fig.~\ref{Upper-near}(a--d)],
which results in larger interaction with the PEC core. Hence, the
RCS reduction in the far field is degraded with the increase of $x$ in the
concave-down class as shown in Fig.~\ref{Upper-far}.

\section{Justification of Improved Segmentation}
It has been pointed out that usually we require a large value of $M$
during the discretization process, otherwise the conversion scheme
in Fig.~\ref{Conversion} will lose its accuracy. We also find that a
nonlinear transformation with a given $x$ of its concave-up class
will yield a better invisibility by taking the segmentation as
Eq.~(\ref{lower-seg}) in the physical space (this is also the only
way to determine sub-layers of type-A and type-B dielectrics in
practice). Here the justification is given. The concave-up class
with $x=4$, which is characterized by the mapping curve
$r'(4)=\frac{R_2R_1^4}{R_2^4-R_1^4}\Big[(r/R_1)^4-1\Big]$, is
considered as an example.

\begin{table}
\caption{The total scattering cross section (unit: dBsm) for
concave-up nonlinearly transformed spherical cloaks ($x=4$) under
various segmentations in physical space. }\label{ta1}
\begin{center}
   \begin{tabular}{c|c|c|c|c|c}  \hline\hline
  $ r_n(x)~\mbox{in~Eq.~(13)} $ & $ r_n(1) $ & $ r_n(2) $ & $ r_n(4) $ & $ r_n(6) $ & $ r_n(10) $
  \\ \hline
  $r'(4)~\mbox{in~Eq.~(11)}$ &  $ -3.3 $ & $ -3.5 $ & $ -18.9 $ & $ -3.8 $ & $ -3.1 $\\
\hline\hline
\end{tabular}\\.
{\small $^*$ Note that $r_n(1)$ for concave-up cloaks corresponds to
the case of equally dividing the physical space.}
\end{center}
\end{table}

In Table I, different segmentation $r_n(x)$ sets (see
Eq.~\ref{lower-seg}) for real radius $r$ are applied to $r'(4)$,
e.g., the sets of $x=1$ ($r_n(1)$), $x=2$ ($r_n(2)$), $x=6$
($r_n(6)$) and $x=10$ ($r_n(10)$). The total cross sections
corresponding to those four sets are compared with the best
segmentation set of its own $r_n(4)$, i.e., equally dividing the
virtual space of $x=4$ into 40 initial-layers and then projecting
onto physical space. We find that, for concave-up class $x=4$,
$r_n(4)$ is indeed its optimal segmentation among these
possibilities.

\section{Conclusion}
In this paper, we discussed two novel classes of
nonlinear-transformation--based spherical cloaks. An approximation
mechanism for such classes of spherical cloaks by concentric
isotropic sub-layers is proposed. The optimal segmentation in the
virtual space is presented, and the role of nonlinearity in the
coordinate transformation is discussed in order to provide better
invisibility performance for the proposed classes of nonlinear
spherical cloaks in both the near
  and far field. We find that concave-up nonlinear-transformation
spherical cloaks exhibit better and more stable invisibility
performance than the corresponding segmentation of
  linear-transformation or concave-down cloaks. An interesting
  topic for future exploration would be to consider a wider class of
  segmentation schemes, for example by using the segmentation
  presented here as the starting point for a numerical optimization (e.g., Ref. \cite{Cummer_PRA} for cylindrical cloaks) of
  variable segment thicknesses and material parameters.

\section{Acknowledgement}
This research was supported in part by the Army Research Office
through the Institute for Soldier Nanotechnologies under Contract
No. W911NF-07-D-0004.

\end{document}